\documentclass{ws-mpla}
\usepackage{psfig}
\begin{document}

\markboth{Saneesh Sebastian and V C Kuriakose}
{Dirac quasinormal modes of MSW black holes }

%%%%%%%%%%%%%%%%%%%%% Publisher's Area please ignore %%%%%%%%%%%%%%
\catchline{}{}{}{}{}
%%%%%%%%%%%%%%%%%%%%%%%%%%%%%%%%%%%%%%%%%%%%%%%%%%%%%%%%%%%%%%%%%%%

\title{Dirac Quasinormal modes of MSW black holes}

\author{\footnotesize  Saneesh Sebastian\footnote{
Typeset names in 8 pt Times Roman, uppercase. Use the footnote to 
indicate the present or permanent address of the author.}}

\address{Department of Physics, Cochin University of Science and
Technology, Kochi 682022, India\\
saneeshphys@cusat.ac.in}

\author{V C Kuriakose}

\address{Department of Physics, Cochin University of Science and
Technology, Kochi 682022, India\\
vck@cusat.ac.in
}

\maketitle

\pub{Received (Day Month Year)}{Revised (Day Month Year)}

\begin{abstract}
In this paper we study the Dirac quasinormal modes of an uncharged $2+1$ black hole proposed by Mandal et. al 
and referred to as MSW black hole in this work. The quasinormal mode is studied using WKB approximation method.
The study shows that the imaginary part of quasinormal frequencies increases 
indicating that the oscillations are damping and hence the black hole is stable against Dirac perturbations.

\end{abstract}
\ccode{PACS Nos.: 04.70.Dy, 04.70.-s }
\section{Introduction}
           The study of quasinormal modes has a long history\cite{kd}. Vishveshwara\cite{vr} first presented the idea of quasinormal modes. Chandrasekhar 
and Detweller\cite{ss} studied the quasinormal modes of Schwarzschild black hole. The study of quasinormal modes started with black hole stability 
problem which concern the evolutions of black hole perturbations. The evolution of wave fields around a black hole consists of roughly
three stages\cite{htc,az,jj}. The first is an initial wave burst coming directly from source and is dependent on the initial form of the original wave 
field. The second part involves damped oscillations called quasinormal frequencies, which is independent of the initial value of the wave 
but depends on the background black hole spacetime. Because of radiation damping, the normal frequencies are complex. The last stage is the 
power law tail which arises because of back scattering of long range gravitational field\cite{vp}.

	  It is believed that the presence of black hole can be inferred using this quasinormal waves. Thus recently 
the study of quasinormal modes becomes important as a number of gravitational wave detectors are expected to start operating soon. Usually the quasinormal mode 
frequencies are damped complex frequencies and thus we must employ the numerical methods to evaluate them. Unlike the case of gravitational, 
electromagnetic and scalar perturbations, the Dirac field is more complicated to evaluate. The third order WKB method\cite{bc,sc,si} is utilized for computation of 
quasinormal modes in the present study.

	  $2+1$ dimensional black holes are the simplest toy model of general $3+1$ black holes in General Theory of Relativity. One 
example of $2+1$ black holes is the BTZ black holes\cite{btz}. The scalar, electromagnetic and gravitational perturbations can be analytically 
studied for such black holes\cite{vc}, but it is difficult in the case of Dirac field. Such studies using WKB method are there in the literature.
Another type of $2+1$ black holes is charged dilaton black holes. Quasi normal modes of charged dilaton black hole is studied\cite{sf}. A special case of such black holes is the MSW black holes.
It is a one parameter family of chargeless  black holes proposed by Mandal et. al\cite{gm}. Previously we have studied the thermodynamics and spectroscopy of MSW black hole\cite{sv} 

	  Here in the present work we study the massless Dirac quasinormal modes of MSW black holes using WKB approximation. The paper is organized
as follows in section. 2 we introduce the MSW black hole. In section. 3 we discuss the Dirac field in MSW space-time background. In section. 4 we
evaluate the quasinormal modes and finally in section. 5 give the conclusions.
\section{The MSW black hole }
The metric for MSW black hole is given by ($c=G=1$ system of units)
\begin{equation}
ds^{2}=-U(r)dt^{2}+\frac{dr^{2}}{U(r)}+\gamma^{2}r d\phi^{2},
\end{equation}
where U(r) is given by\cite{cm}
\begin{equation}
U(r)=8\Lambda\beta r-2M\sqrt{r}.
\end{equation}
This metric has a zero at (ie the horizon) 
\begin{equation}
r_{+}=\frac{M^{2}}{16\Lambda^{2}\beta^{2}},
\end{equation}
 where $\Lambda$ is the cosmological constant, $\beta$ is a constant factor, $M$ is the mass of the black hole. 
\section{Dirac field in MSW space-time background}

\begin{table}[h]
\tbl{Quasinormal frequencies for $\beta=0.1$.}
{\begin{tabular}{@{}cccc@{}} \toprule
k & n &Re(E) &
Im(E) \\
\colrule
1\hphantom{0} & \hphantom{0}0 & \hphantom{0}0.142987 & -0.113155 \\
2\hphantom{0} & \hphantom{0}0 & \hphantom{0}0.279143 & -0.114674 \\
\hphantom{0} & 1& 0.286690\hphantom{0} & -0.341522\hphantom{0} \\
3 & 0 & 0.415674\hphantom{0} & -0.115149\\
\hphantom{0} &1&0.421666\hphantom{0}&-0.343879\\
\hphantom{0} &2&0.431245\hphantom{0}&-0.571293\\
4 & 0 & 0.552724\hphantom{0} & -0.115311\\
\hphantom{0} &1& 0.557569\hphantom{0}&-0.344896\\
\hphantom{0} &2& 0.565582\hphantom{0}&-0.573155\\
\hphantom{0} &3& 0.576227\hphantom{0}&-0.800933\\
5 & 0 & 0.690033\hphantom{0} & -0.115378\\
\hphantom{0} &1& 0.694058\hphantom{0}&-0.345414\\
\hphantom{0} &2& 0.700958\hphantom{0}&-0.574273\\
\hphantom{0} &3& 0.710081\hphantom{0}&-0.802461\\
\hphantom{0} &4& 0.721375\hphantom{0}&-1.030450\\
\botrule
\end{tabular}\label{ta1} }
\end{table}
\begin{figure}[h]
\centering
\includegraphics{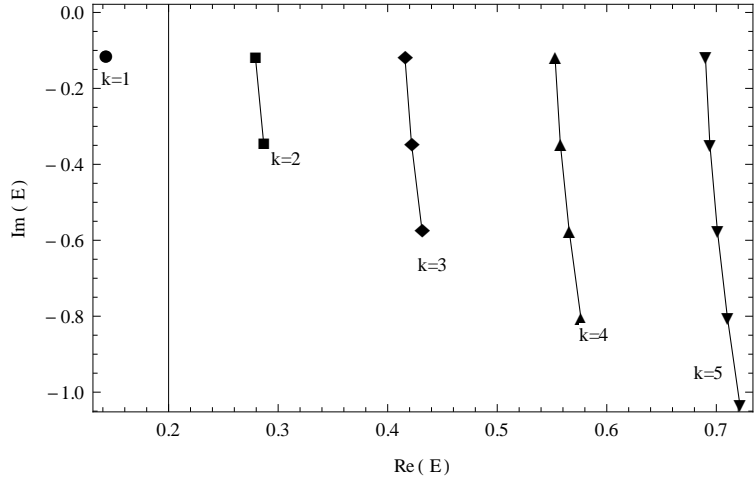}
\caption{Quasinormal frequencies of MSW black hole with $\beta=0.1$}
\end{figure}
In this section we consider the Dirac field in MSW background space-time. The Dirac equation in a general background space-time 
 can be written as\cite{dr} 
\begin{equation}
[i\gamma^{\mu}\partial_{\mu}-i\gamma^{\mu}\Gamma_{\mu}]\Phi=m\Phi,
\end{equation}
where $m$ is the mass of the Dirac field, $\gamma^{\mu}$ are the curvature dependent Dirac matrices and are represented in terms of 
tetrad field as
\begin{equation}
 \gamma^{\mu}=e^{\mu}_{a}\gamma^{a},
\end{equation}
$\gamma^{a}$ represent the standard flat space Dirac matrices, which satisfy 
\begin{equation}
\{\gamma^{a},\gamma^{b}\}=2\eta^{ab}.
\end{equation}
$e^{\mu}_{a}$, the tetrad field is defined by
\begin{equation}
g_{\mu\nu}=\eta^{ab}e^{a}_{\mu}e^{b}_{\nu},
\end{equation}
where $\eta^{ab}=diag(-1,1,1,1)$ being the Minkowski metric. $\Gamma_{\mu}$, the spin connections are given by 
\begin{equation}
\Gamma_{\mu}=\frac{1}{8}[\gamma^{a},\gamma^{b}]e^{\nu}_{a}e_{b\nu;\mu},
\end{equation}
where $e_{b\nu;\mu}=\partial_{\mu}e_{b\nu}-\Gamma^{\alpha}_{\mu\nu}e_{b\alpha}$ is the covarient derivative of $e_{b\nu}$ with
$\Gamma^{\alpha}_{\mu\nu}$ being Christoffel symbols. With the MSW metric given as above we can take the tetrad as  
\begin{eqnarray}
e^{\mu}_{a}=diag\left(-\left(8\Lambda\beta r-2M\sqrt{r}\right)^{-\frac{1}{2}},\left(8\Lambda\beta r-2M\sqrt{r}\right)^{\frac{1}{2}},0,r \right)
\end{eqnarray}
The spin connection satisfy the equation 
\begin{equation}
[\Gamma_{\mu},\gamma^{\nu}]=\frac{\partial \gamma^{\nu}}{\partial x^{\mu}}+\Gamma^{\nu}_{\mu \rho}\gamma^{\rho}.
\end{equation}
We solve equation (10) for spin connection $\Gamma_{\mu}$ and is given by 
\begin{eqnarray}
 \Gamma_{0}=-\frac{1}{2}\left(4\Lambda\beta r-\frac{m}{2\sqrt{r}}\right)\left(\gamma_{1}\gamma_{0}\right), \Gamma_{1}=0,\nonumber\\
  \Gamma_{2}=0, \Gamma_{3}=\frac{1}{2}\left(8\Lambda\beta r-2m\sqrt{r}\right)^{\frac{1}{2}}\left(\gamma_{1}\gamma_{3}\right),
\end{eqnarray}
and hence $\gamma^{\mu}\Gamma_{\mu}$ becomes 

\begin{table}[h]
\tbl{Quasinormal frequencies for $\beta=0.05$.}
{\begin{tabular}{@{}cccc@{}} \toprule
k & n &Re(E) &
Im(E) \\
\colrule
1\hphantom{0} & \hphantom{0}0 & \hphantom{0}0.051064 & -0.056245 \\
2\hphantom{0} & \hphantom{0}0 & \hphantom{0}0.099745 & -0.056991 \\
\hphantom{0} & 1& 0.103979\hphantom{0} & -0.169241\hphantom{0} \\
3 & 0 & 0.147834\hphantom{0} & -0.057383\\
\hphantom{0} &1&0.151505\hphantom{0}&-0.170971\\
\hphantom{0} &2&0.157294\hphantom{0}&-0.284075\\
4 & 0 & 0.196101\hphantom{0} & -0.057550\\
\hphantom{0} &1& 0.199216\hphantom{0}&-0.171805\\
\hphantom{0} &2& 0.204175\hphantom{0}&-0.285420\\
\hphantom{0} &3& 0.210956\hphantom{0}&-0.398999\\
5 & 0 & 0.244513\hphantom{0} & -0.057627\\
\hphantom{0} &1& 0.247177\hphantom{0}&-0.172257\\
\hphantom{0} &2& 0.251510\hphantom{0}&-0.286211\\
\hphantom{0} &3& 0.257315\hphantom{0}&-0.399993\\
\hphantom{0} &4& 0.264698\hphantom{0}&-0.513824\\
\botrule
\end{tabular}\label{ta1} }
\end{table}
\begin{figure}[h]
\centering
\includegraphics{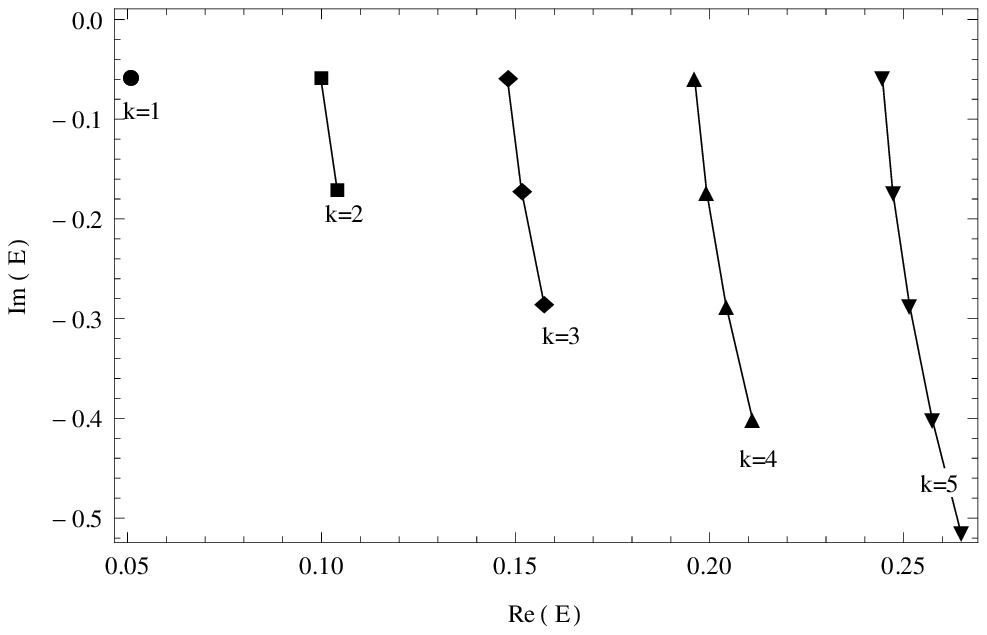}
\caption{Quasinormal frequencies of MSW black hole with $\beta=0.05$}
\end{figure}
 
\begin{table}[h]
\tbl{Quasinormal frequencies for $\beta=0.01$.}
{\begin{tabular}{@{}cccc@{}} \toprule
k & n &Re(E) &
Im(E) \\
\colrule

3 & 0 & 0.0135923\hphantom{0} & -0.0113035\\
\hphantom{0} &1&0.0144644\hphantom{0}&-0.0334106\\
\hphantom{0} &2&0.0158395\hphantom{0}&-0.0555517\\
4 & 0 & 0.0179184\hphantom{0} & -0.0113719\\
\hphantom{0} &1& 0.0187814\hphantom{0}&-0.0337322\\
\hphantom{0} &2& 0.0201505\hphantom{0}&-0.0560795\\
\hphantom{0} &3& 0.0221412\hphantom{0}&-0.0784838\\
5 & 0 & 0.0222176\hphantom{0} & -0.0114231\\
\hphantom{0} &1& 0.0230410\hphantom{0}&-0.0339561\\
\hphantom{0} &2& 0.0243446\hphantom{0}&-0.0564386\\
\hphantom{0} &3& 0.0262203\hphantom{0}&-0.0789680\\
\hphantom{0} &4& 0.0286961\hphantom{0}&-0.1015530\\
\botrule
\end{tabular}\label{ta1} }
\end{table}
\begin{equation}
\gamma^{\mu}\Gamma_{\mu}=\gamma_{1}\left(8\Lambda\beta r-2M\sqrt{r}\right)^{\frac{1}{2}}\left[\frac{1}{2r}+\frac{\left(4\Lambda\beta-\frac{M}{2\sqrt{r}}\right)}{2\left(8\Lambda\beta r-2M\sqrt{r}\right)}\right].
\end{equation}
Therefor the Dirac equation becomes
\begin{equation}
\left[i\gamma_{0}\left(8\Lambda\beta r-2M\sqrt{r}\right)^{\frac{-1}{2}}\partial_{t}+i\gamma_{1}\left(8\Lambda\beta r-2M\sqrt{r}\right)^{\frac{1}{2}}\left[\partial_{r}+\frac{1}{2r}\right]+i\frac{\gamma_{3}}{r}\partial_{\phi}-m\right]\Phi=0.
\end{equation}
This equation can be transformed using tortoise coordinates to Schrodiger like equations as,
\begin{equation}
 \left(-\frac{d^{2}}{dr_{*}^{2}}+V_{(+)1}\right)F^{+}=E^{2}F^{+}
\end{equation}
and
\begin{equation}
 \left(-\frac{d^{2}}{dr_{*}^{2}}+V_{(+)2}\right)G^{+}=E^{2}G^{+},
\end{equation}
where $r_{*}$ is the tortoise coordinate, $F$ and $G$ are the two wave functions of the Dirac equation. 
$V_{(+)1,2}$ are given by 
 
\begin{equation}
V_{(+)1,2}=\pm\frac{dW_{(+)}}{dr_{*}}+W^{2}_{(+)},
\end{equation}
where $W$ is a function of $k$ related to the potential.  
The same set of equations repeat with $(+)$ changed to $(-)$. Putting together we get 
\begin{equation}
 \left(-\frac{d^{2}}{dr_{*}^{2}}+V_{1}\right)F=E^{2}F,
\end{equation}
\begin{equation}
 \left(-\frac{d^{2}}{dr_{*}^{2}}+V_{2}\right)G=E^{2}G.
\end{equation}
The effective potential can be derived from this as 
\begin{equation}
V(r,k)=\left(\frac{\sqrt{\Delta}k}{r^{2}}\right)^{2}+\frac{\Delta}{r^{2}}\frac{d}{dr}\left(\frac{\sqrt{\Delta}k}{r^{2}}\right).
\end{equation}
For our metric given by Eq.($1\&2$) the potential can be written as 
\begin{equation}
V(r,k)=\frac{k^{2}}{r^{2}}\left(8\Lambda\beta r-2M\sqrt{r}\right)+\frac{k}{2r^{2}}\left(8\Lambda\beta r-2M\sqrt{r}\right)^{\frac{1}{2}}\left(-8\Lambda\beta r+3M\sqrt{r}\right).
\end{equation}
\begin{figure}[h]
The potential is plotted for different values of k is given in figure 
\centering
\includegraphics{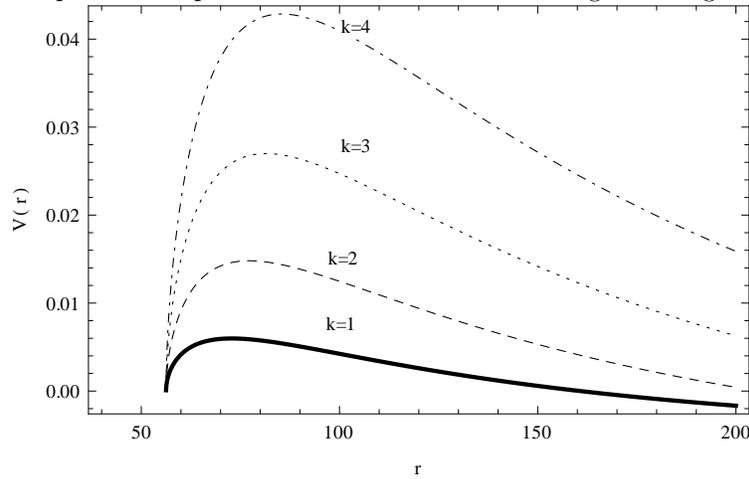}
\caption{Variation of potential with k}
\end{figure}
Fig.(3) shows the dependence of potential on the angular momentum quantum number k. It is of barrier form and the barrier increases in height 
as the k value increases.
\section{Evaluation of Quasinormal Modes for massless Dirac Field}
In order to evaluate the Dirac quasinormal modes we use the WKB approximation method developed by Schutz, Will and Iyer\cite{bc,sc,si}. Comparing with other numerical 
methods this method has been found to be accurate up to around one percent for both real and imaginary parts. The equation for computing the quasinormal modes 
$E$ is given by, 
\begin{equation}
E^{2}=[V_{0}+(-2V''_{0})^{\frac{1}{2}}\Lambda]-i(n+\frac{1}{2})(-2V_{0}'')^{\frac{1}{2}}(1+\Omega),
\end{equation}
where
\begin{equation}
\Lambda=\frac{1}{(-2V_{0}'')^{\frac{1}{2}}}\left[\frac{1}{8}\left(\frac{V_{0}^{(4)}}{V_{0}''}\right)\left(\frac{1}{4}+\alpha^{2}\right)-\frac{1}{288}\left(\frac{V_{0}'''}{V_{0}''}\right)^{2}(7+60\alpha^{2})\right],
\end{equation}
and
\begin{eqnarray}
\Omega=\frac{1}{(-2V_{0}'')} [ \frac{5}{6912}\left(\frac{V_{0}'''}{V_{0}''}\right)^{4}\left(77+188\alpha^{2}\right)
-\frac{1}{384}\left(\frac{(V_{0}''')^{2}(V_{0}^{(4)})}{(V_{0}'')^{3}}\right)\left(51+100\alpha^{2}\right) \\ \nonumber
+\frac{1}{2304}\left(\frac{V_{0}^{(4)}}{V_{0}''}\right)^{2}(67+68\alpha^{2})+
\frac{1}{288}\left(\frac{V_{0}'''V_{0}^{(5)}}{(V_{0}'')^{2}}\right)(19+28\alpha^{2})\\ \nonumber
\frac{1}{288}\left(\frac{V_{0}^{(6)}}{V_{0}''}\right)(5+4\alpha^{2})],
\end{eqnarray}
where $\alpha=n+\frac{1}{2}$, and 
\begin{equation}
V_{0}^{(n)}=\frac{d^{n}V}{dr_{0}^{n}}|_{r_{*}=r_{*}max}.
\end{equation}
Substituting the potential $V(r)$ obtained from (Eq.20) in Eq.(21) we get the complex quasinormal modes for massless Dirac field in MSW space-time. 
The values are listed in tables and plotted. Fig.(1) and Fig.(2) represent the quasinormal mode frequencies for
different values of $\beta$. The figures show that the real part of complex frequencies slightly increase, while the magnitude of imaginary part of
frequencies increases rapidly with mode number $n$ for the same value of $k$. This indicates that the higher modes decay faster than 
the low lying modes. 
\section{Conclusion} 
In this work we have studied the quasinormal modes of a chargeless $2+1$  black hole called MSW black hole. The complex frequencies are 
found out and tabulated. Graphs are plotted for various values of the parameter $\beta$. The modulus value of imaginary part of the frequencies
increases with increasing mode number showing that the radiation is damping which in turn shows that the black hole is stable against massless 
Dirac perturbations. Real values of the frequencies increases in this case contradictory to the normal $3+1$ black holes. Cho\cite{htc} obtained
the Dirac quasinormal modes of Schwarzschild black hole using WKB approximation method.They showed that the real part of quasinormal modes increases
with angular momentum quantum number $k$ and for a particular $k$, the real part of quasinormal modes decreases with increasing $n$. Jiliang\cite{jj}
got the quasinormal modes of Schwarzschild black hole using continued fraction and Hill-determinant methods and obtained an increasing real part. In both cases the imaginary 
part increases with mode number showing damping and thus black holes in $3+1$ dimensions are stable against Dirac perturbation. This result agrees
with the present work for $2+1$ black hole. 
\section*{Acknowledgement}
SS wishes to thank CSIR, New Delhi for financial support under CSIR-SRF scheme. VCK wishes to acknowledge Associateship of IUCAA, Pune, India

\end{document}